%Paper: hep-th/9508044
%From: peeters@insti.physics.sunysb.edu (Bas Peeters)
%Date: Wed, 9 Aug 1995 16:57:40 -0400 (EDT)
%Date (revised): Fri, 11 Aug 1995 12:44:05 -0400 (EDT)

%%%%%%%%%%%%%%%%%%%%%%%%%%%%% harvmac.tex %%%%%%%%%%%%%%%%%%%%%%%%%%%%%%%%%%
\input harvmac.tex
%\draft
\Title{\vbox{\baselineskip12pt\hbox{ITP-SB-95-24}\hbox{hep-th/9508044}}}
{\vbox{\centerline{String Theory near a Conifold Singularity}}}
%\footnote{$^\dagger$}{Work supported in part by NSF grant No. PHY9309888.}
%For more complicated situations, substitute for {\it either\/} argument:
%\Title{\vbox{\baselineskip12pt\hbox{HUTP-88/A000}\hbox{SLAC-PUB 88-001}
%               \hbox{photocopy at own risk}}}
%{\vbox{\centerline{This title is too long to fit}
%       \vskip2pt\centerline{comfortably on one line*}}}
%   \footnote{}{*optional footnote on title}

\centerline{ Dileep P. Jatkar\footnote{$^\dagger$}{Work supported in part
by NSF grant No. PHY9309888. Address after Oct. 1st : Mehta Research
Institute of Mathematics \& Mathematical Physics, 10 Kasturba Gandhi Marg,
Allahabad 211002, India.} and Bas Peeters\footnote{$^\star$}{Address after
Sept. 20th : Queen Mary and Westfield College, London E1 4NS, England.}}
\smallskip\centerline{\it Institute for Theoretical Physics}
\centerline{\it State University of New York at Stony Brook}
\centerline{\it Stony Brook, NY 11794-3840, USA}

\vskip .3in
We demonstrate that type II string theory compactified on a singular
Calabi-Yau manifold is related to $c=1$ string theory compactified at
the self-dual radius. We establish this result in two ways. First we
show that complex structure deformations of the conifold correspond,
on the mirror manifold, to the problem of maps from two dimensional
surfaces to $S^2$. Using two dimensional QCD we show that this problem
is identical to $c=1$ string theory. We then give an alternative
derivation of this correspondence by mapping the theory of complex
structure deformations of the conifold to Chern-Simons theory on
$S^3$. These results, in conjunction with similar results obtained for
the compactification of the heterotic string on $K_3\times T^2$,
provide strong evidence in favour of S-duality between type II strings
compactified on a Calabi-Yau manifold and the heterotic string on
$K_3\times T^2$.

\Date{08/95}

\lref\godI{E. Witten, {\it String theory dynamics in various dimensions},
hep-th/9503124.}

\lref\godII{E. Witten, Nucl. Phys. {\bf B373} (1992) 187.}

\lref\godIII{E. Witten, {\it Chern-Simon gauge theory as a string theory},
hep-th/9207094.}

\lref\godIV{E. Witten, Comm. Math. Phys. {\bf 121} (1989) 351.}

\lref\godV{E. Witten, Nucl. Phys. {\bf B371} (1992) 191.}

\lref\sen{A. Sen, {\it String-string duality conjecture in six dimensions and
charged solitonic strings}, hep-th/9504027.}

\lref\harstro{J. A. Harvey and A. Strominger, {\it The heterotic string is a
soliton}, hep-th/9504047.}

\lref\kava{S. Kachru and C. Vafa, {\it Exact results for N=2
compactifications of heterotic strings}, hep-th/9505105.}

\lref\strom{A. Strominger, {\it Massless black holes and conifolds in string
theory}, hep-th/9504090.}

\lref\vafa{C. Vafa, {\it A stringy test of the fate of the conifold},
hep-th/9405023.}

\lref\ghova{D. Ghoshal and C. Vafa, {\it $c=1$ string theory as the topological
theory of the conifold}, hep-th/9506122.}

\lref\gms{B. Greene, D. Morrison and A. Strominger, {\it Black hole
condensation and the unification of string vacua}, hep-th/9504145.}

\lref\canoss{P. Candelas and X. C. de la Ossa, Nucl. Phys. {\bf B342}
(1990) 246; P. Candelas, P. Green and T. H\" ubsch, Nucl. Phys. {\bf B330}
(1990) 49.}

\lref\agnt{I. Antoniadis, E. Gava, K. S. Narain and T. R. Taylor, {\it N=2
type II-heterotic duality and higher derivative F-terms}, hep-th/9507115.}

\lref\ghomu{D. Ghoshal and S. Mukhi, Nucl. Phys. {\bf B425} (1994) 173.}

\lref\haop{A. Hanany, Y. Oz and R. Plesser, Nucl. Phys. {\bf B425} (1994) 150.}

\lref\ghm{D. Ghoshal and S. Mukhi, {\it Landau-Ginzburg model for a critical
topological string}, as referred to in \refs{\ghova}.}

\lref\sendual{A. Sen, {\it Strong-weak coupling duality in the four
dimensional heterotic string theory}, hep-th/9402002}

\lref\huto{C. Hull and P. Townsend, {\it Unity of superstring dualities},
hep-th/9410167.}

\lref\gm{D. Gross and A. Matytsin, {\it Some properties of large $N$
two dimensional Yang-Mills theory}, hep-th/9410054.}

\lref\doug{M. Douglas, {\it Conformal field theory techniques in large $N$
Yang-Mills theory}, hep-th/9311130.}

\lref\mp{J. Minahan and A. Polychronakos, Phys. Lett. {\bf 312B} (1993) 155.}

\lref\dcmp{A. D'Adda, M Caselle, L. Magnea and S. Panzeri, {\it Two dimensional
QCD on the sphere and on the cylinder}, hep-th/9309107.}

\lref\viper{V. Periwal, {\it Topological closed-string interpretation of
Chern-Simons theory}, hep-th/9305115.}

\lref\kleb{I. Klebanov, {\it String theory in two dimensions},
hep-th/9108019, and references therein.}

\lref\cdl{S. Chaudhuri, H. Dykstra and J. Lykken,
Mod. Phys. Lett. {\bf A6} (1991) 1665.}

\lref\bcov{M. Bershadsky, S. Cecotti, H. Ooguri and C. Vafa,
Comm. Math. Phys. {\bf 165} (1994) 311.}

\lref\ground{D. Ghoshal, P. Lakdawala and S. Mukhi, Mod. Phys. Lett. {\bf A8}
(1993) 3187; J. Barbon, Int. J. Mod. Phys. {\bf A7} (1992) 7579;
S. Kachru, Mod. Phys. Lett. {\bf A7} (1992) 1419.}

\lref\thooft{G. 't Hooft, Nucl. Phys. {\bf B72} (1974) 461.}

\lref\migd{A.A. Migdal, Sov. Phys. JETP {\bf 42} (1975) 413.}

\lref\muva{S. Mukhi and C. Vafa, {\it Two dimensional black-hole as a
topological coset model of c=1 string theory}, hep-th/9301083.}

\lref\daje{S. R. Das and A. Jevicki, Mod. Phys. Lett. {\bf A5} (1990) 1639.}

\lref\erik{E. Verlinde, Nucl. Phys. {\bf B300} (1988) 360.}

\lref\kac{V. Ka\v c and M. Wakimoto, Adv. Math. {\bf 70} (1988) 156.}

\lref\clz{M. Camperi, F. Levstein and G. Zemba, Phys. Lett. {\bf B247} (1990)
549.}

%\newsec{Introduction}

\noindent{\bf 1.} Considerable progress has been made in the recent past
towards understanding the non-perturbative aspects of string
theory. One of the interesting proposals\refs{\godI}, see also
\refs{\huto}, is that various different string theories are just
different manifestations of presumably a single big picture. Some
evidence in this direction has been established using the conjectured
string-string duality in six dimensions\refs{\sen,\harstro} and
strong-weak coupling duality (S-duality)\refs{\sendual}. More recently,
Kachru and Vafa\refs{\kava} have given some explicit examples of dual
pairs of string theories with $N=2$ spacetime supersymmetry. They
showed that certain Calabi-Yau compactifications of type II string
theory are related under S-duality to heterotic string theory
compactified on $K_3\times T^2$.

It has been known for quite some time\refs{\canoss} that the leading
terms in the expansion of the prepotential for type II strings
compactified on a Calabi-Yau manifold exhibit logarithmic
singularities. Strominger\refs{\strom} has suggested a beautiful
interpretation of these singularities. He proposed that conifold
singularities in a Calabi-Yau manifold can be understood in terms of
extremal black holes becoming massless. Taking these massless fields
into account, the transition across the boundary between the moduli
space of topologically distinct Calabi-Yau manifolds becomes
smooth\refs{\gms}. As a result different classical vacua of string
theory form a connected web. This may eventually help shed some light
on understanding the non-perturbative ground state of string
theory. It is therefore important to understand the universal
features of the conifold singularity in Calabi-Yau manifolds.

It was shown by Vafa\refs{\vafa} that the one loop free energy of the
effective field theory of type II string near a conifold has
universal behaviour and is given by
\eqn\torus{F_1 = -{1\over 12}\log\mu,}
where $\mu$ is the mass of the extremal black hole which becomes
massless at the conifold singularity. Following this one loop
computation, it was suggested by Ghoshal and Vafa\refs{\ghova} that
the conifold singularity in Calabi-Yau threefolds indeed has universal
behaviour and is given by $c=1$ string theory at the self-dual
radius. They observe that the leading order behaviour (up to two loops
in the string coupling constant) of the free energy of the conifold
theory matches with the perturbative expansion of the free energy of
the $c=1$ string at the self-dual radius.

Antoniadis et al. in their recent work\refs{\agnt} have shown that a
one loop computation in heterotic string theory, compactified near an
enhanced symmetry point in the $K_3\times T^2$ moduli space,
reproduces the full $c=1$ string free energy. This result is
consistent with the conjectured duality between type II and heterotic
strings. To put this conjecture on a more firm footing we need to
better understand the relation between type II strings near a conifold
and $c=1$ string theory.

In this paper we will show that the result of \refs{\ghova} can be
extended to all orders in string perturbation theory using two
different methods.
\item{i)} We employ mirror symmetry to map the type B topological
sigma model corresponding to the string theory near the conifold to
the A-model on the mirror Calabi-Yau. On this mirror manifold the singularity
is resolved into an $S^2$, and in order to study the universal
behaviour of the string near the singularity we only need to consider
mappings from the worldsheet to this $S^2$. This is exactly the
subject of 2D QCD, and using the relation between 2D QCD and $c=1$
string theory\refs{\gm,\doug,\mp,\dcmp} we find the complete
perturbative expansion of the free energy of the conifold theory.
\item{ii)} The theory of complex structure deformations of Calabi-Yau
manifolds is given by Kodaira-Spencer theory which was used in
\refs{\ghova} to study deformations of the conifold singularity. We show
that Kodaira-Spencer theory in the neighbourhood of a conifold
singularity reduces to Chern-Simons theory on $S^3$. This, combined
with the results obtained by Periwal\refs{\viper}, who showed that in
the double scaling limit the free energy of Chern-Simons theory on
$S^3$ is the generating function of the Euler characteristics of
the moduli spaces of surfaces of genus $g$, establishes yet another
relation between the conifold theory and the $c=1$ string.

\medskip
\noindent{\bf 2.} Let us briefly review some results obtained earlier
on the compactification of type II string theory on the
conifold\refs{\strom,\vafa,\ghova}. The conifold singularity occurs in
a Calabi-Yau manifold whenever either a 2-cycle or a 3-cycle shrinks
to zero size\refs{\canoss}. This singularity can therefore be resolved
by replacing it by either an $S^2$ (called small resolution) or an
$S^3$ (deformation). As mentioned earlier, the leading term in the
free energy expansion of the string compactified on a Calabi-Yau
manifold has logarithmic singularities, which can be interpreted as
extremal black holes becoming massless\refs{\strom}. The procedure
followed in \refs{\ghova} to desingularize the conifold involves
deforming the singularity into an $S^3$. Near this singularity the
degenerating Calabi-Yau can be described by
\eqn\conic{x_1^2 + x_2^2 + x_3^2 +x_4^2 = \mu,}
where $x_i$ are coordinates on ${\bf C}^4$. As $\mu\rightarrow 0$, the
quadric develops a singularity. Nonzero $\mu$ corresponds to resolving
this singularity into an $S^3$. Under a linear redefinition, \conic\
possesses a formal similarity with the deformed ground ring of
$c=1$ string theory at the self-dual radius\refs{\godII,\ground}.
The topological Landau-Ginzburg model corresponding to $c=1$ string
theory is given by the singular superpotential\refs{\ghomu,\haop}
\eqn\sup{W(x) = - \mu x^{-1}.}
The central charge of this model is equal to that of a Calabi-Yau
manifold ($\hat c = 3$). By adding quadratic terms to the
superpotential, which does not alter the central charge, we can embed
this superpotential in an ambient space such that geometrically it is
equivalent to a Calabi-Yau\refs{\ghm}. The modified superpotential is
\eqn\msp{W(x) = - \mu x^{-1} + x^2_1 + x^2_2 + x^2_3 + x^2_4.}
The $W=0$ locus of \msp\ satisfies the Calabi-Yau condition and
corresponds to the deformed conifold.

The $S^3$ deformation of the conical singularity corresponds to
complex structure deformations of the Calabi-Yau and hence is
described by a topological sigma model of type B.  Following the
techniques developed in \refs{\bcov}, Ghoshal and Vafa obtain the
terms through genus two in the perturbative expansion of the free
energy of the string theory near a conifold singularity. They show the
equivalence of these terms with the free energy expansion of the $c=1$
string.

Using these ingredients they show that the behaviour of type II
strings near the conifold singularity in a Calabi-Yau manifold is
universal and is captured uniquely by $c=1$ string theory at the
self-dual radius.

\medskip
\noindent{\bf 3.} Mirror symmetry tells us that the topological sigma
model of type B on a Calabi-Yau manifold is identical to a topological sigma
model of type A on the corresponding mirror manifold. Here we consider
the A-model on the mirror, which is the appropriate model to study
when we choose to resolve the conifold singularity into an $S^2$.

Since we are interested in understanding the behaviour of string
theory near the conifold, the only thing that is relevant to
understanding its behaviour near the singularity is string dynamics on
the new $S^2$ generated by the resolution. String theory near the
singularity therefore reduces to the topological sigma model which
counts the instanton maps from the worldsheet to $S^2$. The maps from
two dimensional surfaces to $S^2$ have been studied in the context of
two dimensional QCD. In particular it has been demonstrated that the
latter theory in the $1/N$ expansion can be interpreted in terms of
sums over maps between two dimensional manifolds, with the target
manifold being the manifold on which the original QCD theory is
defined.

Here we will show, following Gross and Matytsin\refs{\gm}, that two
dimensional QCD on a cylinder in the $1/N$ expansion gives rise to the
free energy of the $c=1$ string at the self-dual radius. From the
above discussion one would infer that the two sphere, instead of the
cylinder, is the space on which the QCD should be defined. This
situation is quite reminiscent of the relation between matrix models
and gauged WZNW models. In this case, the Penner model is related to
the $SU(2)/U(1)$ gauged WZNW model at level $k$, which is analytically
continued to $k=-3$\refs{\godV}. It can also be thought of as an
$SL(2)/U(1)$ model with $k=3$\refs{\muva}. This analytic continuation
turns the compact coset manifold $SU(2)/U(1)$ into a non-compact one
corresponding to $SL(2)/U(1)$. It would be interesting to understand
this phenomenon at a more fundamental level in this context; however,
we will proceed here with the cylinder as the target space of the two
dimensional QCD.

The partition function for QCD with gauge group $U(N)$ on a cylinder
can be computed exactly for any finite $N$ using heat kernel methods
and equals\refs{\migd}
\eqn\qcdpf{Z_N[U_{C_1},U_{C_2};A] = \sum_R \chi_R(U_{C_1})
\chi_R(U_{C_2}^\dagger) e^{-{\lambda A\over 2N} C_2(R)},}
where $\lambda=g^2N$ is the coupling constant and $A$ is the area of
the cylinder. Since the partition function only depends on their
product, we will set $\lambda=1$. $\chi_R(U)$ is the character of the
matrix $U$ in the representation $R$; $C_2(R)$ is the second
Casimir of this representation.

We now introduce the collective variables
$\sigma(\theta)$ to denote the eigenvalue distribution of the unitary
matrix $U$ in the large $N$ limit. In this collective field theory
approach one can then derive the following equation for the classical
action
\eqn\act{{\partial S\over \partial A} = {1\over 2}
\int\limits_0^{2\pi}\! d\theta \, \sigma_1(\theta) \left[ \left(
{\partial\over \partial\theta} {\delta S\over\delta \sigma_1(\theta)}
\right)^2 - {\pi^2\over 3} \sigma_1^2(\theta) \right].}
This equation, when we interpret $A$ as time, $\sigma_1(\theta)$ as
canonical coordinate, and $\Pi(\theta)$ as its conjugate momentum,
corresponds to the Hamilton-Jacobi equation for the Das-Jevicki
Hamiltonian\refs{\daje}
\eqn\dash{H[\sigma(\theta),\Pi(\theta)] = {1\over 2} \int\limits_0^{2\pi}
\! d\theta \, \sigma(\theta) \left[ \left( {\partial\Pi(\theta) \over
\partial\theta}\right)^2 - {\pi^2\over 3} \sigma^2(\theta) \right].}
This Hamiltonian describes the effective field theory of $c=1$ string
theory. The free energy of this theory is given by\refs{\kleb,\cdl}
\eqn\fe{F[\mu, R]={1\over 2} \mu^2 \log\mu - {1\over 24}\left( R + {1\over
R}\right) \log\mu + \sum_{g\ge 2} F_g(R) \mu^{2-2g},} where $R$ is the
radius of compactification and $\mu$ is the renormalized string
coupling constant. The functions $F_g$ are invariant under
$R\leftrightarrow 1/R$ and for $R=1$ equal the Euler character of the
moduli space of genus $g$ Riemann surfaces. Matching this expression
with the result of the one loop computation\refs{\vafa} in the
conifold theory given in \torus , we find that the conifold
corresponds to $c=1$ string compactified at the self-dual radius
($R=1$). Hence the expansion of the free energy of the conifold around
$\mu=0$ is equal to the topological expansion of the $c=1$ string.

\medskip
\noindent{\bf 4.}
Complex structure deformations of a Calabi-Yau manifold are described
by Kodaira-Spencer theory. This theory has been shown to reduce to
Chern-Simons theory\refs{\bcov} in the following situation. Consider
analytic continuation of a Calabi-Yau manifold to a six dimensional
symplectic manifold which consists of a 3-dimensional base space $X$
and a 3-dimensional internal space $Y$. The holomorphic and the
anti-holomorphic 3-form inherited from the Calabi-Yau manifold become
volume forms on X and Y. The action of the Kodaira-Spencer theory then
coincides with the action of the Chern-Simons theory on $X$ with the
infinite dimensional group of volume preserving diffeomorphisms of $Y$
as its gauge symmetry. Thus for a fixed holomorphic 3-form $\Omega$ on
a Calabi-Yau manifold, which becomes the volume form of $Y$, the
Kodaira-Spencer action reduces to the Chern-Simons action.

The $1/N$ expansion\refs{\thooft} of Feynman graphs in the
Chern-Simons theory\refs{\godIII} is a topological expansion with
$N^2$ being a genus counting parameter. Periwal has shown\refs{\viper}
that the exact free energy of $SU(N)$ Chern-Simons theory at level $k$
can be expanded in powers of $(N+k)^{-2}$. Due to the level-rank
duality symmetry of this expansion it is possible to do both strong
coupling ($N$ large with $k$ fixed) and weak coupling ($k$ large with
$N$ fixed) expansions.

Let us come back to the conifold theory. In case of the B-model the
conifold singularity is deformed into an $S^3$. Since the base of the
cone is $S^3\times S^2$, near the apex of the cone the deformed
geometry is like $S^3\times R^3$. This can be interpreted as a
6-dimensional symplectic manifold with the symplectic structure
obtained from the K\" ahler structure of the Calabi-Yau manifold. In
the neighbourhood of this $S^3$ Kodaira-Spencer theory reduces to
Chern-Simons theory with $\Omega$-preserving diffeomorphism
symmetry. The partition function of the Chern-Simons theory on
$S^3$\refs{\godIV} is given by
\eqn\csp{Z[S^3]=S_{0,0},}
where $S_{\alpha}^{\beta}$ is the modular transformation matrix
corresponding to the action of $SL(2,Z)$ on the characters of the
$SU(N)$ WZNW model at level $k$. The partition function for gauge
group $SU(N)$ therefore can be written as\refs{\erik,\kac,\clz}
\eqn\pf{Z[S^3; N, k]=(N+k)^{-N/2}\sqrt{{N+k\over N}}\prod_{j=1}^{N-1}\left[
2\sin\left({j\pi\over N+k}\right)\right]^{N-j}.}  In the limit
$N\rightarrow\infty$, this Chern-Simons theory possesses volume
preserving symmetry. Type II string near a conifold singularity is in
the strong coupling regime and since the Chern-Simons theory is an
appropriate description near a conifold, the strong coupling expansion
($N$ large with $k$ finite) of its free energy should capture the
behaviour of string theory near a conifold. Let us define a new
variable $x=N/(N+k)$. Then in the scaling limit where $x\rightarrow 1$
and $k=(N+k)(1-x)$ is held fixed, expansion of the free energy of the
Chern-Simons theory on $S^3$ is given by
\eqn\euler{F[S^3; k]= {1\over 2}k^2 \log k -{1\over 12} \log k +
\sum_{g\ge 2} \chi_g k^{2-2g} + \hbox{{\rm terms analytic in }} k,}
where $\chi_g=B_{2g}/2g(2g-2)$ is the Euler character of the moduli
space of Riemann surfaces of genus $g$\refs{\viper}. After identifying
$k$ with the cosmological constant $\mu$ it is easy to recognize that
this is the well known free energy expansion of $c=1$ string
theory. Hence we see that the universal behaviour of the conifold
theory is given by $c=1$ string theory.

\medskip
\noindent{\bf 5.} In this paper we established the relation
between the conifold theory and $c=1$ string theory using two
techniques. We first showed that complex structure deformations of
the conifold singularity correspond to maps from 2D surfaces to $S^2$
in the mirror manifold. With the aid of 2D QCD we demonstrated the
equivalence of the free energies of the conifold theory and $c=1$
string theory. We then rederived this relation by studying
Chern-Simons theory on $S^3$. Our results along with the results of
\refs{\agnt} on the heterotic side present a strong case in favour of
S-duality in four dimensional $N=2$ string theories.

We would like to end with an interesting observation. As is well
known, $c=1$ string theory can be represented in terms of free
fermions. In this free fermion representation the $\mu\rightarrow 0$
limit corresponds to vanishing Fermi energy and therefore this theory
has gapless excitations. It would be interesting to see if these
excitations have any relation to the extremal black holes that become
massless at the conifold singularity.

\bigskip
\noindent{\bf Acknowledgements:} We would like to thank N. Berkovits,
D. Ghoshal, R. Gopakumar, M. Ro\v cek, W. Siegel and C. Vafa for
valuable discussions.
%\vfil\eject
\listrefs
\bye